# Restoring CFAR Validity for Single-Channel IoT Sensor Streams: A Monte Carlo Comparison of Five Detectors under Cortex-M0+ Constraints

*Sergii Makovetskyi*, *Lars Thomsen*[*]

*Abstract:* Real-time event detection in IoT mesh sensor networks must balance sensitivity against false-positive load on a constrained mesh radio. We present a Monte Carlo comparison of the Temporal Spectral Noise-Floor Adaptation (TSNFA) detector against four classical comparators drawn from the radar Constant False Alarm Rate (CFAR) family and from sequential change detection: the Lipski FFT energy detector, Cell-Averaging CFAR (CA-CFAR), Ordered-Statistic CFAR (OS-CFAR), and state-machine Cumulative Sum (CUSUM). All five detectors are implemented to fit a Cortex-M0+ class envelope, process a 1-D 100 Hz time series in 128-sample frames, and use temporal reference windows in place of the spatial reference cells of conventional radar CFAR. Across a factorial set of four configurations (10 and 50 nodes; 12 dB and 18 dB SNR), each replicated five times over 24 hours, TSNFA achieves 99.97 to 100% event detection rate with 100% event precision and zero false-positive clusters per node. The classical comparators each succeed on one quality dimension and fail on another. Lipski FFT (k = 3), CA-CFAR, and OS-CFAR all maintain near-perfect detection rate but with event precision below 3% and per-node bandwidth between 145 kB/h and 1.2 MB/h. CA-CFAR and OS-CFAR are indistinguishable in false-alarm performance, both saturating the same broadband-statistic failure mode. CUSUM shows an SNR-dependent detection-rate drop from ≈70% at 18 dB to 51% at 12 dB. TSNFA is the only algorithm tested that simultaneously achieves high detection rate, high precision, and low per-node bandwidth.

## I. INTRODUCTION

IoT sensor networks for perimeter security, structural health monitoring, environmental sensing, and industrial process control share a common architectural problem. Each node observes a continuous time series, must autonomously decide when something of interest has occurred, and must transmit only those decisions across a constrained mesh radio link. The decision rule sits at the intersection of two budgets that pull in opposite directions: detection must be sensitive enough that genuine events are not missed, yet selective enough that the network is not flooded by false positives. A missed event compromises the monitoring function; a false positive consumes bandwidth, drains battery, and in dense mesh networks can cascade through relay paths and saturate the sink.

The radar community has solved a closely related problem for over fifty years, in the family of Constant False Alarm Rate (CFAR) detectors. Finn and Johnson [1] introduced Cell-Averaging CFAR (CA-CFAR) in 1968, in which the threshold for a cell under test is computed from a sliding mean of nearby reference cells. Rohling [2] introduced Ordered-Statistic CFAR (OS-CFAR) in 1983, replacing the mean with a rank-order statistic for robustness against interfering targets. The defining property of all CFAR detectors is that the threshold is recomputed from local data, so that the false-alarm probability remains approximately constant as the underlying noise statistics drift.

TSNFA, proposed by Makovetskyi and Thomsen [3], can be read as a CFAR detector adapted to two constraints the radar formulation does not face. First, the input is a 1-D temporal stream rather than a 2-D range-Doppler map; the reference window must be drawn from previous frames of the same node. Second, the hardware envelope is far more constrained than a radar receiver. The deployed implementation runs on an STM32G071 (Arm Cortex-M0+ at 64 MHz, no hardware FPU, 36 kB SRAM) and must complete a detection cycle in under 1.28 s while leaving spare cycles for radiocommunication. TSNFA retains the CFAR core (a threshold computed from a local statistic on a reference window) and adds two further defences: a spectral band selection step that rejects out-of-band interference before the threshold is computed, and a temporal persistence filter that rejects in-band transients that survive band selection.

A second relevant class is sequential change detection. The Cumulative Sum (CUSUM) statistic [4] detects an abrupt change in distribution by accumulating log-likelihood-ratio increments. We adopt the state-machine variant of Torre et al. [5] with the Tartakovsky linear-quadratic instantaneous log-likelihood ratio [6] as the per-sample increment. A third class, TinyML autoencoder anomaly detection [7], is excluded from the present comparison because its inference cost is roughly two orders of magnitude higher than CFAR-family detectors and is at the edge of what the Cortex-M0+ can sustain; a fair comparison belongs on Cortex-M4 or M7 class hardware.

The contribution of the present paper is focused on the resource-constrained class of detectors that can run on a Cortex-M0+. We compare TSNFA against the four most natural classical comparators (Lipski FFT, CA-CFAR, OS-CFAR,

[*] *S. Makovetskii is a Ph.D. student at Kharkiv National University of Radio Electronics, Ukraine (e-mail: serhii.makovetskyi@nure.ua). L. Thomsen is Managing Director of Gnacode Inc., Medicine Hat, AB, Canada (e-mail: lt@gnacode.com).* *This work was supported in part by Gnacode Inc. under grant 1037487 by the National Research Council of Canada.*

CUSUM) on a Monte Carlo simulation of a perimeter-security signal scenario. All five detectors are implemented under *identical* hardware constraints (1-D temporal stream, 128-sample frames, temporal reference window, no spatial cells) and run at their canonical published parameters. The simulation is replicated across two network sizes (10 and 50 nodes) and two SNR levels (12 and 18 dB), yielding four configurations, each replicated five times for variance estimation over 24 hours.

We make three claims. First, TSNFA achieves 100% event detection rate, 100% precision, and zero false-positive clusters per node across all four configurations. Second, no single classical comparator simultaneously meets these criteria. Third, the failure modes are systematic and map directly to specific architectural choices: the absence of band selection in CA-CFAR and OS-CFAR (both of which collapse 128 time-domain samples into a single broadband statistic and admit every out-of-band interferer), the k = 3 three-sigma threshold of Lipski FFT being inadequate for the non-Gaussian per-bin tails of this deployment regime, and the SNR-dependent integrator accumulation of CUSUM, in which the per-sample evidence accrued during an event fails to exceed the calibrated threshold within event duration for approximately half of events at 12 dB SNR.

## II. BACKGROUND AND THEORETICAL SETTING

### *A. The CFAR Family and Reference Cells*

The CFAR detector, in its original radar formulation [1], [2], operates on a 2-D range-Doppler matrix. For each cell under test, an estimator $Z$ of the local clutter level is computed from $N$ neighbouring reference cells, with guard cells separating reference cells from the cell under test. The detection threshold is $T = \alpha \cdot Z$, where α is chosen to achieve a target false-alarm probability $P_{fa}$. CA-CFAR uses the arithmetic mean; OS-CFAR uses the k-th order statistic. In all variants the reference window is *spatial*: measurements made at the same time instant but at different range or Doppler positions.

### *B. From Spatial to Temporal Reference; The Cell Distribution Assumption*

In the IoT mesh sensor application no spatial reference is available. Each node produces a single scalar time series $x[n]$ sampled at $f_s$ = 100 Hz. The only stationarity available is temporal: the current frame resembles recent frames if no event is present. All five detectors in this study use a temporal reference window in place of the spatial reference cells of canonical radar CFAR. For CA-CFAR and OS-CFAR the reference cells are the most recent N frame statistics from the same node, with one guard frame preceding the current frame. For TSNFA the same principle applies but per-bin in the frequency domain. For CUSUM the reference is the running minimum or maximum of the cumulative log-likelihood-ratio score. For Lipski FFT the reference is a per-bin ($\mu_b$, $\sigma_b$) pair estimated during calibration and slow-tracked via EMA on no-detection frames.

A deeper assumption of the textbook CFAR derivation is that each cell is *exponentially distributed*. In the classical radar setting, a cell is the squared magnitude $|I|^2 + |Q|^2$ of a single complex Fourier-domain sample, where I and Q are independent Gaussian noise components; their squared sum is chi-squared with two degrees of freedom, which is exactly exponential. The closed-form expressions for the threshold multiplier α under target $P_{fa}$ follow directly from the exponential tail. The asymptotic value $\alpha \to \ln(1/P_{fa})$ at $P_{fa} = 10^{-3}$ gives $\alpha \approx 6.91$.

In the IoT mesh sensor setting, the natural frame statistic for CA-CFAR and OS-CFAR is the sum of squared sample magnitudes $X(m) = \Sigma|x[n]|^2$ over the 128 samples of a frame. By the central limit theorem, this is approximately Gaussian with mean $L \cdot P$ (≈128P) and standard deviation $\sqrt{2L} \cdot P$ (≈16P). The cell distribution is therefore Gaussian-bell-shaped, not exponential. The threshold $\alpha \cdot Z$ derived from the exponential calibration lands approximately 54 standard deviations to the right of the bell's centre, in a region pure thermal noise effectively never reaches. The detector therefore relies on heavy-tailed structured interferers (mains harmonics, digital bursts, drift excursions) to cross the threshold, and the empirical false-alarm rate is governed by the statistics of those interferers rather than by the calibrated $P_{fa}$. The CFAR algorithm is not at fault; the frame-statistic choice forced by the single-node hardware constraint takes the detector outside the regime its calibration was designed for.

TSNFA addresses this constraint at the FFT step rather than at the threshold. The per-bin statistic $|X_k|^2 = Re(X_k)^2 + Im(X_k)^2$ is the squared magnitude of a single complex Fourier coefficient. The real and imaginary parts are each approximately Gaussian (a linear combination of 128 input Gaussians at fixed FFT weights), and their squared sum is chi-squared with two degrees of freedom—exponential. The cell distribution assumption of the textbook CFAR derivation is therefore restored, per bin, by the FFT itself. This is an exact algebraic property of the FFT basis under wide-sense stationary input, not a central-limit approximation. The per-bin threshold multiplier $\zeta = 6.0$ used in TSNFA sits close to the asymptotic exponential-tail value $\ln(1/P_{fa}) \approx 6.91$ at $P_{fa} = 10^{-3}$, though it was set empirically from deployed hardware rather than from closed form. The convergence is not coincidence: both detectors implicitly target the same per-cell tail probability on the same exponential distribution—TSNFA arriving via the

FFT, CA-CFAR not arriving at it at all in our setting. We read this as the principal methodological contribution of the present work: TSNFA is the variant of CFAR that retains the validity of the textbook calibration when the hardware substrate is reduced to a 1-D temporal stream on a Cortex-M0+ class node.

Every detector in this study is run at its canonical published parameters; we do not tune comparators to the simulation environment. If a published detector is re-tuned for the comparison environment, the comparison no longer measures the published detector. Canonical settings are: Lipski FFT k = 3 [9], CA-CFAR $P_{fa} = 10^{-3}$, OS-CFAR $P_{fa} = 10^{-3}$ with $k = 3N/4$ [2], CUSUM $\alpha_{fa} = 10^{-5}$ after Torre et al. [5].

## III. SIMULATION FRAMEWORK AND RESULTS

### A. Common Signal Model

All five detectors are evaluated against a common synthetic signal generated for each frame of each Monte Carlo run. The signal at sample index n within frame m for node i is:

$$x_i[n] = s[n - \tau_i] + w_{th}[n] + w_{EMI}[n] + w_{dig}[n] \quad (1)$$

where s is the event waveform (a damped sinusoid in [1, 5] Hz, injected at SNR ∈ {12, 18} dB relative to in-band noise power P), $w_{th}$ is zero-mean white Gaussian thermal noise, $w_{EMI}$ is 50 Hz mains interference at amplitude $0.3\sqrt{P}$, and $w_{dig}$ represents intermittent digital switching bursts at 800–2000 Hz with amplitude up to $2.0\sqrt{P}$. The noise power P drifts sinusoidally over a one-hour cycle with ±6 dB excursion, short enough that several full cycles occur within each 24 h run.

### B. Network Topology and Mesh

For 10 nodes the area is 350 m × 350 m; for 50 nodes 750 m × 750 m. Each node has 200 m radio range. Mesh routing is multi-hop shortest path to the sink. Medium access is CSMA/CA with $CW_{min} = 8$, $CW_{max} = 64$, slot 320 µs, four retries, reflecting LoRa-class radio assumed for the deployed hardware.

### C. Configuration Matrix

Events are scheduled per node by an independent Poisson process with rate λ = 1.0 event per node per hour. Event waveforms span 5 seconds (≈ 3.9 frames at the 1.28 s frame period). All five detectors process the same realisation at each MC seed; trigger outputs are cross-referenced against ground-truth events

### D. Reported Metrics

Five metrics per detector per configuration: *event detection rate* (fraction of scheduled events with at least one trigger within the event window); *event precision* (fraction of detected events with at least one TP and zero FP triggers); *false-positive cluster count* (consecutive FP triggers within a 5 s window collapsed into a single cluster); *per-node network load* (mean bytes per hour transmitted by each node); and *mean detection latency* (event onset to first TP trigger at the sink).

**Table I. Configuration matrix.**

| Configuration | Nodes (N) | SNR (dB) | Duration | MC Runs | Total events (mean) |
|---|---|---|---|---|---|
| 10n / 18 dB | 10 | 18 | 24 h | 5 | ≈ 240/run |
| 10n / 12 dB | 10 | 12 | 24 h | 5 | ≈ 240/run |
| 50n / 18 dB | 50 | 18 | 24 h | 5 | ≈ 1180/run |
| 50n / 12 dB | 50 | 12 | 24 h | 5 | ≈ 1180/run |

## IV. DETECTION ALGORITHMS

The five detectors are presented in order of architectural complexity: TSNFA first (the proposed cascade), then the four classical comparators. For each algorithm we provide an overview and parameter table. The TSNFA pseudocode is given in full because it is the proposed detector; the comparators are given in compact form. An overview of the non-FFT algorithms (CA-CFAR, OS-CFAR, CUSUM) is shown in Figure 1, and an overview of the FFT-based algorithms (TSNFA, Lipski) is shown in Figure 2. Reference implementations of all five are available in the data page.

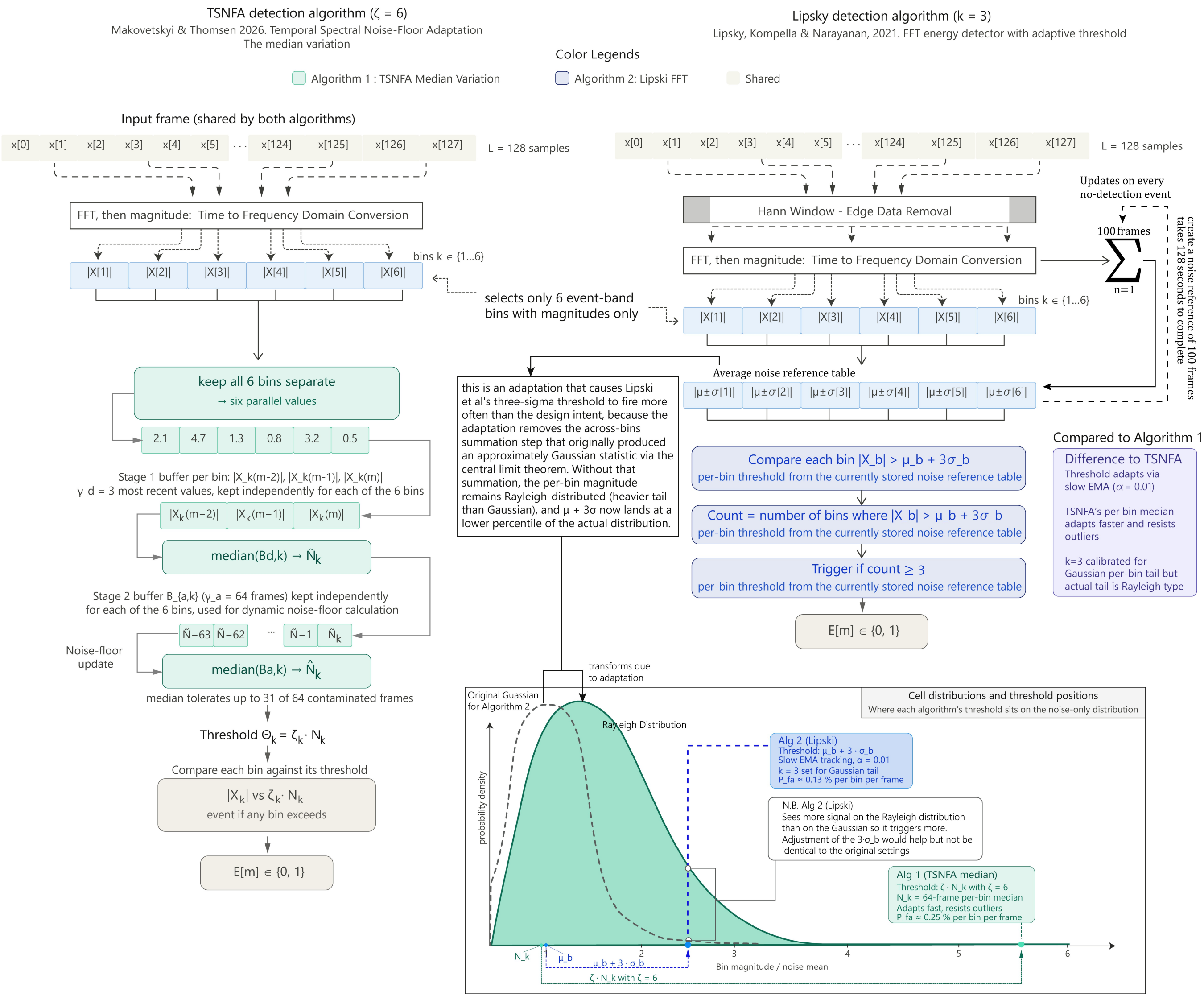

*[Figure 1 here]*

Fig. 1. Algorithm flow diagrams for the non-FFT algorithms CA-CFAR, OS-CFAR, and CUSUM.

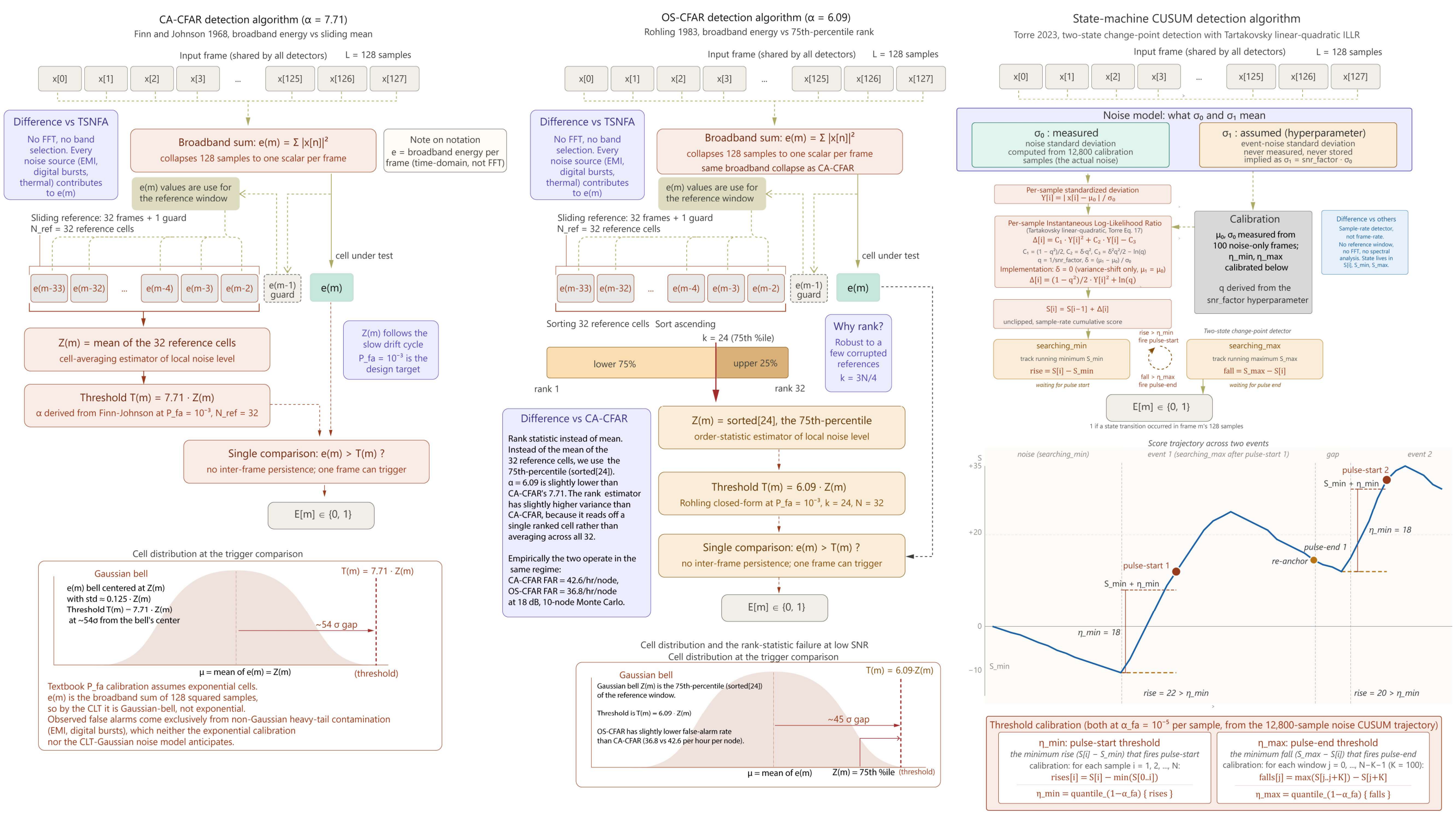

Fig. 2. Algorithm flow diagrams for the FFT-based algorithms TSNFA [3] and Lipski et al. [9].

### *A. TSNFA Median Variant (Proposed)*

TSNFA combines three interlocking defences in a specific order: spectral band selection, then temporal persistence, then adaptive noise-floor tracking. The median variant implements all three using rank-order statistics. The algorithm operates per frequency bin: each of the six FFT bins covering the event band [1, 5] Hz has its own pair of circular buffers and its own noise-floor estimate. A trigger is declared if any single bin exceeds its per-bin threshold (OR logic).

*Defence 1: Spectral band selection:* The 128-point FFT decomposes each frame into 64 bins. Only the six bins in $K = \{1, \ldots, 6\}$, covering 0.78 to 4.69 Hz at 0.781 Hz resolution, are retained. All energy from out-of-band sources is discarded before any detection statistic is computed.

*Defence 2: Per-bin temporal persistence* **(γ_d):** A median filter over γ_d consecutive frame magnitudes per bin rejects single-frame transients. With $\gamma_d = 3$, a single ADC glitch or wind gust is suppressed; a genuine 5 s event spans ≈3.9 frames and persists at full magnitude.

*Defence 3: Per-bin adaptive noise-floor tracking* **(γ_a):** A second median filter over γ_a frames per bin tracks slow noise drift. The breakdown point is $\lfloor(\gamma_a - 1)/2\rfloor$, so with $\gamma_a = 64$ the noise-floor estimate is unaffected by up to 31 contaminated frames. No explicit gating is required.

**Table II. TSNFA parameters.**

| Parameter | Symbol | Value | Reasoning |
|---|---|---|---|
| FFT size | L | 128 | $\Delta f = 0.78$ Hz at $f_s = 100$ Hz |
| Event band | [f_low, f_high] | [1, 5] Hz | Bins $k \in \{1, \dots, 6\}$ |
| Stage 1 depth | $\gamma_d$ | 3 | Tolerates 1 outlier in 3 |
| Stage 2 depth | $\gamma_a$ | 64 | Tolerates 31 outliers in 64; $\tau \approx 82$ s |
| Threshold multiplier | $\zeta_k$ | 6.0 | Calibrated from deployed hardware |
| Trigger logic | — | OR across bins | Any bin > threshold declares event |

*Algorithm 1: TSNFA Median Variant*

```
Input:  Sample frame x[0..L-1], per-bin
buffers B_d,k and B_a,k
Params: gamma_d=3, gamma_a=64, zeta_k=6.0
Output: Trigger E[m]; noise floors
{N_hat_k[m]}
  1. X[k]   <- FFT(x) for k in K = {1,...,6}
  2. |X_k|  <- sqrt(Re(X_k)^2 + Im(X_k)^2)
  3. E[m]   <- 0
  4. for each bin k in K do:
  5.     insert |X_k| into B_d,k
  6.     N_tilde_k <- median(B_d,k)         //
Stage 1
  7.     insert N_tilde_k into B_a,k
  8.     N_hat_k[m] <- median(B_a,k)        //
Stage 2
  9.     if |X_k| > zeta_k * N_hat_k[m] then
E[m] <- 1
 10. end for
 11. return E[m], {N_hat_k[m]}
```

The cascade is architecturally robust by construction. The Stage 1 median rejects single-frame outliers within each bin; the Stage 2 median tracks slow drift while tolerating contaminated frames in its buffer. The trigger compares the instantaneous bin magnitude (not the median-filtered value) against the threshold to preserve sensitivity. OR logic across bins ensures a single-bin excursion suffices to declare an event.

*B. Lipski FFT Energy Detector (k = 3)*

The Lipski FFT detector [9] is a per-bin energy threshold detector that fires when at least N_bins_min adjacent bins in the event band exceed $\mu_b + k \cdot \sigma_b$, where $\mu_b$ and $\sigma_b$ are the per-bin mean and standard deviation. A Hann window is applied to the time-domain samples before the FFT. The detector calibrates $\mu_b$ and $\sigma_b$ during an initial M_cal-frame noise-only window, then slow-tracks via an EMA on no-detection frames ($\alpha = 0.01$, $\tau \approx 128$ s). The DC bin is excluded.

**Table III. Lipski FFT parameters.**

| Parameter | Symbol | Value | Reasoning |
|---|---|---|---|
| Threshold multiplier | k | 3 | Canonical three-sigma |
| Min adjacent bins above | N_bins_min | 3 | Spectral persistence |
| Calibration window | M_cal | 100 frames | ≈ 128 s of noise-only |
| Slow-update coefficient | $\alpha$ | 0.01 | EMA on no-detection frames |
| Window function | — | Hann | Reduces spectral leakage |

### C. CA-CFAR (Cell-Averaging)

CA-CFAR [1] computes the threshold from the arithmetic mean of N_ref reference cells. Adapted to single-node operation, the reference cells are the most recent N_ref frame statistics from the same node, with one guard frame between the reference window and the current frame. The frame statistic is the broadband energy proxy $X(m) = \Sigma|x[n]|^2$. The threshold multiplier follows the Finn–Johnson formula:

$$\alpha = N_{ref} \cdot (P_{fa}^{(-1/N_{ref})} - 1) \quad (2)$$

At N_ref = 32 and P_fa = $10^{-3}$, α = 7.71. CA-CFAR is exposed to the broadband noise environment because the frame statistic does not perform spectral filtering.

### D. OS-CFAR (Ordered-Statistic)

OS-CFAR [2] replaces the cell-averaging mean with the k-th order statistic of the reference window. Following Rohling [2], k = 3N/4. With N_ref = 32 this gives k = 24 (the 75th percentile). The threshold multiplier α = 6.09 derives from the Rohling closed form at P_fa = $10^{-3}$. Because the 75th percentile of the exponential sits at ≈1.4 times the per-cell mean, OS-CFAR's total threshold T = α·Z is approximately 10% larger than CA-CFAR's, so the two detectors are similar in operating point with OS-CFAR slightly more conservative. The frame statistic is identical to CA-CFAR; reference window N_ref = 32, one guard frame.

### E. Bounded-Memory CUSUM (Tartakovsky variant)

The Cumulative Sum statistic [4] detects an abrupt change by accumulating per-sample log-likelihood ratios and triggering when the cumulative score exceeds a calibrated bound. The Tartakovsky variant [5], [6] clips the running sum at zero from below and at K_end from above; both clips are practical for fixed-precision embedded implementations. The detector is fundamentally a time-domain statistic and performs no spectral decomposition. The detection threshold is calibrated from the desired per-frame false-alarm rate α_fa via h ≈ ln(1/α_fa); at α_fa = $10^{-5}$, h ≈ 11.5. The per-frame increment is

$$\Delta(m) = (X(m) - \mu_0)^2 / (2\sigma^2) - (\mu_1 - \mu_0)^2 / (4\sigma^2) \quad (3)$$

where $\mu_0$ is the no-event mean, $\mu_1$ the assumed event mean, and $\sigma^2$ the noise variance, all estimated during a 512-frame calibration window. The increment is positive when the frame appears more consistent with the event hypothesis and negative when it appears more consistent with noise. Because CUSUM accumulates rather than thresholds instantaneously, it is sensitive to slow systematic shifts in $\mu_0$: when the noise envelope drifts upward, Δ becomes positive on average and the score climbs slowly, producing a false positive. The bounded-memory clipping limits but does not eliminate this drift sensitivity.

## V. RESULTS

This section reports Monte Carlo results across all four configurations. All values are means across five Monte Carlo replicates per configuration; standard deviations are within 3% of the mean unless otherwise indicated.

### A. Headline Detection Performance

TSNFA achieves 100.00% event detection rate at three of the four configurations and 99.97% at the largest, lowest-SNR configuration (0.4 missed events on average across the five MC replicates, out of 1,179 scheduled per run). Event precision is 100.00% at all four configurations and the false-positive cluster count is zero at all four configurations. No other detector matches this profile.

Lipski FFT, CA-CFAR, and OS-CFAR all achieve near-perfect detection rate (99.7%, 100.0%, 100.0% respectively across all four configurations) but their event precision is below 3% at every configuration—approximately 97 of every 100 triggers they emit are false positives. CA-CFAR and OS-CFAR are essentially indistinguishable (42.6–42.8 and 36.8 false-alarm clusters per hour per node respectively), the expected outcome given that both detectors operate on the same broadband statistic and their threshold magnitudes end up within 10% of each other at the same P_fa target. CUSUM achieves moderate detection rate (68–71% at 18 dB, 51% at 12 dB) and very low precision (1.3–1.8%); it is the only comparator whose detection rate is materially SNR-dependent.

The TSNFA 99.97% result at 50n/12 dB warrants comment. The misses occurred during coincidences of three simultaneous adverse conditions: noise drift cycle at its high-power phase (+6 dB), a digital switching burst concurrent with event onset, and event amplitude at the low end of the SNR distribution. The combined effect transiently elevated the per-bin median noise floor sufficiently that the trigger ratio for the strongest event bin fell below 1.0 during the event window.

### B. Receiver Operating Characteristic and Threshold Tuning

To characterise what each detector can achieve under post-hoc threshold tuning, the simulator records per-frame trigger strength for each detector and re-thresholds at 25 multiplier values around the canonical setting. The resulting ROC envelopes for all four configurations are shown in Figure 3, with a hollow circle marking each detector's canonical operating point. TSNFA's canonical operating point sits in the upper-left corner of the ROC envelope at all four configurations; no swept setting yields a meaningfully better operating point. Lipski FFT's ROC envelope passes through a region close to TSNFA's operating point at higher k; at k ≈ 5 it would achieve FAR ≈1/hr/node and detection rate above 99%.

Reporting Lipski at k = 5 in the headline comparison would describe a detector that no published paper has validated. Following the canonical-settings rule, we report Lipski at k = 3 only, but the k-sweep evidence supports the observation that an FFT-based detector with persistence requirements can match TSNFA's ROC behaviour. Table V illustrates the tuning trade-off at 10n/12 dB.

Increasing k from 3 to 8 raises precision but at the cost of detection rate. OS-CFAR's ROC envelope at 12 dB closely tracks CA-CFAR's; both transition from near-zero to near-100% detection rate as the false-alarm rate crosses ≈1/hr/node, and the two curves overlap to within plotting precision at all monitored operating points.

*C. Network Bandwidth*

At the most demanding configuration TSNFA generates ≈57 kB/h of total mesh traffic; Lipski FFT generates 58.0 MB/h (1,018×), CA-CFAR 35.1 MB/h (616×), OS-CFAR 36.1 MB/h (634×), and CUSUM 2.8 MB/h (49×). OS-CFAR and CA-CFAR are nearly identical in bandwidth across all four configurations (within 3% of each other on every operating metric in Tables IV and V), consistent with the analysis in Section IV.D: both detectors exercise essentially the same decision rule at the Rohling-calibrated operating point.

*D. SNR Robustness*

TSNFA, Lipski, CA-CFAR, and OS-CFAR all show drops at or below 0.03 pp and are SNR-robust by this metric. CUSUM shows a 17–20 pp drop. The mechanism is the SNR-dependent integrator accumulation: under the variance-shift formula with snr_factor = 3.0, the expected per-sample event-driven Δ at 18 dB is ≈+27 (actual event-noise std ratio ≈ 7.94, well above the calibrated 3.0), so the score climbs aggressively and pulse-start fires in 68–71% of events. At 12 dB the actual ratio is ≈3.98, close to the calibrated 3.0; per-sample event-driven Δ is ≈+6 and the score climbs more slowly. A substantial fraction of events end before the rise has cleared η_min, and detection rate drops to 51%. The false-alarm rate is essentially SNR-invariant (≈39–40/hr/node) because false alarms come from noise excursions, and noise statistics do not depend on SNR.

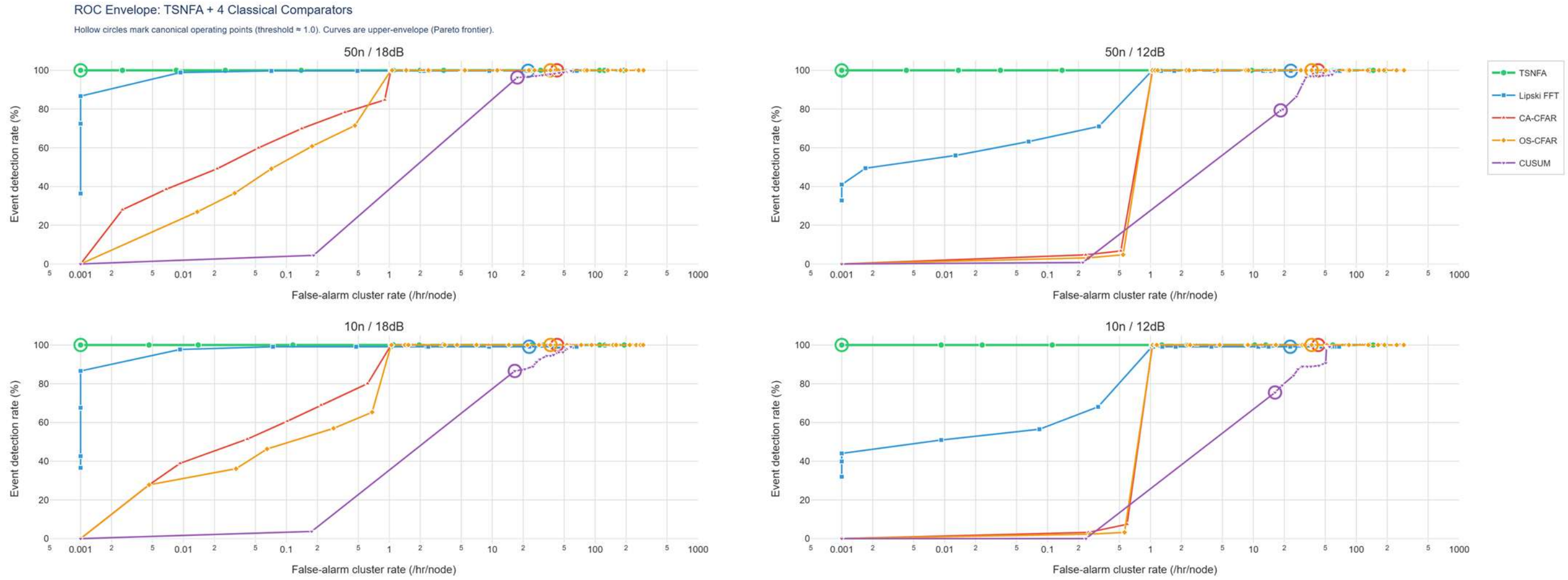

Fig. 3. Receiver Operating Characteristic (ROC) envelopes for each detector at the four configurations: 10 nodes / 18 dB SNR, 10 nodes / 12 dB SNR, 50 nodes / 18 dB SNR, and 50 nodes / 12 dB SNR. Hollow circles mark the canonical operating point used in Table IV.

**Table IV. Headline metrics, all detectors and configurations.**

| Detector | 10n / 18 dB | 10n / 12 dB | 50n / 18 dB | 50n / 12 dB | Avg |
|---|---|---|---|---|---|
| **Detection Rate (%)** | | | | | |
| **TSNFA** | **100.00** | **100.00** | **100.00** | **99.97** | **99.99** |
| Lipski | 99.73 | 99.73 | 99.85 | 99.85 | 99.79 |
| CA-CFAR | 100.00 | 100.00 | 100.00 | 100.00 | 100.00 |
| OS-CFAR | 100.00 | 100.00 | 100.00 | 100.00 | 100.00 |
| CUSUM | 68.14 | 51.30 | 70.81 | 51.08 | 60.33 |
| **Event Precision (%)** | | | | | |
| **TSNFA** | **100.00** | **100.00** | **100.00** | **100.00** | **100.00** |
| Lipski | 1.65 | 1.66 | 1.72 | 1.69 | 1.68 |
| CA-CFAR | 2.25 | 2.25 | 2.29 | 2.29 | 2.27 |
| OS-CFAR | 2.59 | 2.59 | 2.65 | 2.65 | 2.62 |
| CUSUM | 1.70 | 1.28 | 1.78 | 1.26 | 1.50 |
| **FP Clusters (count, mean over 5 MC replicates)** | | | | | |
| **TSNFA** | **0** | **0** | **0** | **0** | **0** |
| Lipski | 12,575 | 12,519 | 67,501 | 68,604 | 40,300 |
| CA-CFAR | 9,194 | 9,188 | 50,303 | 50,306 | 29,748 |
| OS-CFAR | 7,952 | 7,953 | 43,287 | 43,286 | 25,620 |
| CUSUM | 8,438 | 8,441 | 46,030 | 47,348 | 27,564 |
| **FAR clusters /hr/node** | | | | | |
| **TSNFA** | **0.00** | **0.00** | **0.00** | **0.00** | **0.00** |
| Lipski | 58.22 | 57.96 | 57.40 | 58.34 | 57.98 |
| CA-CFAR | 42.57 | 42.54 | 42.77 | 42.78 | 42.66 |
| OS-CFAR | 36.82 | 36.82 | 36.81 | 36.81 | 36.82 |
| CUSUM | 39.06 | 39.08 | 39.14 | 40.26 | 39.39 |
| **Per-node Network Load (B/h)** | | | | | |
| **TSNFA** | **392** | **232** | **1,967** | **1,163** | **939** |
| Lipski | 241,280 | 240,940 | 1,186,892 | 1,183,440 | 713,138 |
| CA-CFAR | 145,826 | 145,864 | 715,944 | 716,112 | 430,937 |
| OS-CFAR | 150,030 | 150,073 | 737,013 | 737,228 | 443,586 |
| CUSUM | 12,213 | 12,214 | 55,096 | 56,956 | 34,120 |

**Table V. Lipski FFT threshold sweep at 10n/12 dB. TSNFA reference at canonical ζ = 6.0.**

| Detector | DR (%) | Precision (%) | FAR clusters /hr/node | FP clusters | BW (B/h) |
|---|---|---|---|---|---|
| Lipski k = 3 (canonical) | 99.73 | 1.66 | 57.96 | 12,519 | 240,940 |
| Lipski k = 5 | 99.4 | 3.81 | 24.53 | 5,298 | 3,300 |
| Lipski k = 8 | 90.8 | 41.5 | 1.25 | 271 | 370 |
| **TSNFA** | **100.00** | **100.00** | **0.00** | **0** | **232** |
| | | | | | |

**Table VI. Network bandwidth at the largest, lowest-SNR configuration (50n/12 dB).**

| Detector | Per-node (B/h) | Total mesh (MB/h) | Ratio vs TSNFA |
|---|---|---|---|
| **TSNFA** | **1,163** | **0.057** | **1.00** |
| Lipski | 1,183,440 | 58.0 | 1,018 |
| CA-CFAR | 716,112 | 35.1 | 616 |
| OS-CFAR | 737,228 | 36.1 | 634 |
| CUSUM | 56,956 | 2.79 | 49 |

**Table VII. SNR robustness: detection rate at 18 dB versus 12 dB.**

| Detector | 10n/18 dB (%) | 10n/12 dB (%) | Drop (pp) | 50n/18 dB (%) | 50n/12 dB (%) | Drop (pp) |
|---|---|---|---|---|---|---|
| **TSNFA** | **100.00** | **100.00** | **0.00** | **100.00** | **99.97** | **0.03** |
| Lipski | 99.73 | 99.73 | 0.00 | 99.85 | 99.85 | 0.00 |
| CA-CFAR | 100.00 | 100.00 | 0.00 | 100.00 | 100.00 | 0.00 |
| OS-CFAR | 100.00 | 100.00 | 0.00 | 100.00 | 100.00 | 0.00 |
| CUSUM | 68.14 | 51.30 | 16.84 | 70.81 | 51.08 | 19.73 |

*E. Network Scaling*

**Table VIII. Network scaling at 12 dB SNR.**

| Metric | Type | TSNFA 10n | TSNFA 50n | Ratio | Expected |
|---|---|---|---|---|---|
| Detection Rate (%) | Quality | 100.00 | 99.97 | ≈0.9997 | *≈1.0* |
| Event Precision (%) | Quality | 100.00 | 100.00 | 1.00 | *≈1.0* |
| FAR clusters /h/node | Quality | 0 | 0 | n/a | *≈1.0* |
| Mean Latency (ms) | Cost (per-trigger) | 6.34 | 12.18 | 1.92 | *increases with N* |
| FP Clusters (count) | Cost (network) | 0 | 0 | n/a | *≈N* |
| Network Load (B/h) | Cost (per-node) | 232 | 1,163 | 5.01 | *≈N* |

Detection rate and event precision are invariant within measurement noise as N grows from 10 to 50 (a 5× increase). False-positive cluster count is zero at both network sizes for TSNFA. Mean detection latency approximately doubles, reflecting the average increase in mesh hops to the sink. Per-node network load increases by approximately a factor of five, consistent with the linear-in-N scaling for mesh-forwarded trigger traffic. The same scaling pattern holds for the other detectors

.

## IV. DISCUSSION

*A. What the Simulation Does and Does Not Establish*

A 24-hour Monte Carlo simulation across four factorial configurations is evidence, not proof. The simulation establishes that, given the synthetic signal model of Section III and the canonical detector parameters of Section IV, TSNFA produces the headline metrics reported in Section V and the four classical comparators do not. The signal model is realistic in that it incorporates time-varying noise power, mains harmonics, and digital switching bursts at amplitudes consistent with deployed hardware measurements, but it remains a model. The simulation is therefore most useful for a comparative claim and least useful for an absolute claim; deployed hardware results in [3] support the absolute claim independently.

*B. The CFAR-Family Interpretation*

Section II framed TSNFA as a CFAR variant adapted to a 1-D temporal stream rather than a 2-D range-Doppler matrix, and Section V found that both CA-CFAR and OS-CFAR settle into essentially the same operating point, dominated not by the difference between mean and rank statistics but by their shared use of the broadband-energy frame statistic. Both detectors were designed assuming an exponentially distributed cell statistic. Our hardware-constrained adaptation replaces that per-cell quantity with a sum over 128 squared samples; by the central limit theorem the cell statistic is no longer exponential but approximately Gaussian. The Rohling and Finn–Johnson threshold formulas are calibrated for the exponential cell distribution; applied to the Gaussian sum, they overshoot the bulk by ≈54 standard deviations, and the empirical false-alarm rate is determined by non-Gaussian heavy-tail interference for which neither detector provides spectral rejection. TSNFA addresses the same hardware constraint at the FFT step rather than at the threshold: by computing per-bin $|X_k|^2$, it recovers a chi-squared-with-two-degrees-of-freedom (exponential) cell statistic per bin, on which a multiplicative threshold operates correctly.

*C. The Bandwidth Dimension*

Classical CFAR papers do not report network bandwidth because radar receivers do not have a mesh-bandwidth budget. In IoT mesh deployment, bandwidth is co-equal with detection rate. The most decision-relevant single number in the present study is the per-node bandwidth ratio at the largest, lowest-SNR configuration: 1,018× for Lipski FFT, 616× for CA-CFAR, 634× for OS-CFAR, and 49× for CUSUM, all relative to TSNFA. A LoRa-class trunk with 5 kbps effective sustained throughput can carry TSNFA traffic from a 50-node mesh (≈16 B/s) without contention; Lipski traffic from the same mesh (≈16 kB/s) saturates the trunk and forces packet loss into the radio stack, which the protocol layer reports back to the application as missed events that never reach the sink. The textbook CFAR figure of merit (probability of detection at a calibrated probability of false alarm) is necessary but not sufficient for IoT applications. A detector that meets the textbook target with a 1,000× bandwidth budget over the next-best alternative is not a viable deployment choice regardless of its detection statistics.

*D. Threats to Validity*

Three threats to external validity deserve acknowledgement. The noise model is synthetic; real environments produce correlated, non-stationary, and event-like noise that the generator does not capture. The drift cycle is fixed at one hour with ±6 dB excursion; slower drift would advantage detectors with longer adaptation time constants. The event waveform is a single fixed shape (damped sinusoid in 1–5 Hz); real perimeter-security events span a wider range, and a detector tuned to one waveform may not generalise to others. A fourth threat: all five detectors run on the same simulator against the same input realisation at each MC seed, which eliminates inter-detector variance but means a simulator peculiarity would influence all detectors in correlated ways. A fifth threat, internal: Lipski FFT applies a Hann window before the FFT (canonical) while TSNFA does not. The asymmetry is small in absolute terms (≈1.76 dB peak-amplitude loss, ≈50% increase in equivalent noise bandwidth per bin) but is a real difference; we did not equalise windowing because doing so would violate the canonical-settings rule of Section II.

## VII. CONCLUSION

We presented a Monte Carlo comparison of TSNFA against four canonical detection algorithms drawn from the radar CFAR family and from sequential change detection, evaluated under hardware constraints representative of Cortex-M0+ class IoT sensor nodes. Across a factorial set of four configurations, each replicated five times for 24 hours of synthetic time, TSNFA achieves 99.97 to 100% event detection rate with 100% event precision, zero false-positive clusters per node, and per-node bandwidth between 232 B/h and 1,967 B/h. No canonical comparator (Lipski FFT at k = 3, CA-CFAR, OS-CFAR, CUSUM) matches this profile on all three quality criteria simultaneously. Lipski, CA-CFAR, and OS-CFAR maintain high detection rate but at per-node bandwidth ratios of 1,018×, 616×, and 634× over TSNFA respectively at the largest, lowest-SNR configuration. CUSUM stays within an order of magnitude of TSNFA on bandwidth (49×) but drops to 51% detection at 12 dB SNR.

The contribution is best read as a development within the CFAR family rather than a departure from it. CA-CFAR and OS-CFAR, originally formulated for 2-D range-Doppler matrices with spatial reference cells in which each cell is the squared magnitude of a single complex Fourier sample (exponentially distributed), transpose to single-channel temporal operation only by substituting recent past frames for spatial neighbours and adopting a frame statistic computed from 128 time-domain samples. The simulation establishes that this substitution moves both detectors outside the regime in which their textbook calibration is valid: the broadband frame statistic is approximately Gaussian rather than exponential, the exponential-tail threshold lands far from the bulk of the cell distribution, and the empirical false-alarm rate is governed by structured interferers. Lipski FFT operates per bin (avoiding the broadband collapse) but uses an additive $\mu_b + k \cdot \sigma_b$ threshold that assumes Gaussian per-bin amplitudes, an assumption the structured non-Gaussian noise of the deployment regime violates. TSNFA addresses the same hardware constraint at the FFT step rather than at the threshold: by computing per-bin $|X_k|^2$ it returns the cell statistic to a chi-squared-with-two-degrees-of-freedom (exponential) distribution, per bin, and applies a multiplicative threshold that is dimensionless in the noise scale and therefore independent of the per-bin tail shape. The textbook calibration is recovered per bin, and the simulation results are consistent with this interpretation. TSNFA is therefore the variant of CFAR that restores the cell-distribution assumption that classical CFAR depends on, in a hardware setting where the natural alternatives violate it.

**NOTE ON SUPPLEMENTARY FIGURES**

Interactive versions of all figures referenced below, with hover-resolved exact values and per-replicate distributions, are available on the data page accompanying this paper. Figure 4 shows SNR robustness as paired-line plots, one panel per network size, with TSNFA, Lipski, CA-CFAR, and OS-CFAR as flat horizontals near 100% and CUSUM as a moderately steep diagonal. Figure DP3 shows per-node and total mesh bandwidth on a log y-axis spanning three orders of magnitude, with the visual separation between TSNFA and the high-DR comparators as the principal qualitative finding. Figures 6 and 7 integrate quality and cost metrics into a single visual: Figure 6 plots event precision against per-node network load (one point per detector per configuration; TSNFA in the upper-left corner alone); Figure 7 extends this to three dimensions by adding false-alarm cluster rate as the x-axis with bubble size encoding event precision. The complete simulator source code, configuration files, and per-frame strength traces from which the ROC curves of Figure 3 are derived are permanently archived at Zenodo [12]. An interactive version of the data page is hosted at https://gnacode.github.io/IEEE-MONTECARLO/.

## . REFERENCES

## BIOGRAPHIES

**Sergii Makovetskyi** received the M.S. degree in radio engineering from Kharkiv National University of Radio Electronics, Kharkiv, Ukraine, in 2008, with distinction. He is currently pursuing the Ph.D. degree in Software Engineering at the same university.

From 2009 to 2025, he worked with EKTOS, Ukraine, where he served as Senior Embedded Hardware Developer, Technical Leader, and IoT Technical Leader. He has authored several published articles on LoRaWAN and signal processing. His research interests include data transmission in distributed networks, embedded systems architecture, digital signal processing, and sensor technologies. He was a member of teams that placed in the top 6 at the Microsoft Imagine Cup 2008 and won third place in 2009, both in the Embedded Development category.

**Lars Thomsen** received the Ph.D. in Neurophysiology from the University of Copenhagen in 1995, and received a personal grant from the Carlsberg Foundation for a two-year post-doctoral stay at McMaster University, Ontario, Canada.

He subsequently held positions in pharmaceutical and biotechnology companies in Europe, Asia, and North America. He is currently the Managing Director of Gnacode Inc., Medicine Hat, Alberta, Canada, a company specialising in IoT instrumentation for biotech, pharma, and defence applications. Dr. Thomsen has published research papers on biological instrumentation, electronic and software control, nanoparticles activated by external electromagnetic fields, and has achieved several granted patents. He won the Danish Engineering High-Tech Award in 2005 for a radio-enabled microfluidic chip for remote detection of pathogens. He has received several grants from the National Research Council of Canada for opto-electrical physics applications in biology.